\documentclass[article,twocolumn,prl,superscriptaddress,tightenlines]{revtex4}

\usepackage{amssymb}
\usepackage{enumerate}
\usepackage{amssymb}
\usepackage{amsmath}
\usepackage{feynmf}
\usepackage{slashed}
\usepackage{natbib}
\usepackage{graphicx}
\usepackage[pdftex,bookmarks,linktocpage,pdfpagelabels,plainpages=false,hyperfigures,linkcolor=blue,citecolor=blue]{hyperref} 
\hypersetup{colorlinks=true}

\usepackage{mathrsfs,amssymb}  
\usepackage{cancel}
\usepackage[normalem]{ulem}

\newcommand\be{\begin{equation}}
\newcommand\ee{\end{equation}}

\newcommand\bea{\begin{eqnarray}}
\newcommand\eea{\end{eqnarray}}

\newcommand\nn{ \nonumber \\}

\begin{document}
\begin{flushleft}
MI-TH-1925
\end{flushleft}
\bibliographystyle{apsrev4-1}

\title{Bounds on Cosmic Ray-Boosted Dark Matter in Simplified Models
and its Corresponding Neutrino-Floor}

\author{James B.~Dent} 
\affiliation{Department of Physics, Sam Houston State University, Huntsville, TX 77341, USA}

\author{Bhaskar Dutta}
\affiliation{Mitchell Institute for Fundamental Physics and Astronomy,
   Department of Physics and Astronomy, Texas A\&M University, College Station, TX 77845, USA}

\author{Jayden L.~Newstead}
\affiliation{Department of Physics, Arizona State University, Tempe, AZ 85287, USA}

\author{Ian M. Shoemaker}
\affiliation{Center for Neutrino Physics, Department of Physics, Virginia Tech University, Blacksburg, VA 24601, USA}

\begin{abstract}

We study direct detection bounds on cosmic ray-upscattered dark matter in simplified models including light mediators. We find that the energy dependence in the scattering cross section is significant, and produces stronger bounds than previously found (which assumed constant cross sections) by many orders of magnitude at low DM mass. Finally, we compute the ``neutrino-floor'' that will limit future direct detection searches for cosmic ray-upscattered dark matter. While we focus on vector interactions for illustration, we emphasize that the energy dependence is critical in determining accurate bounds on any particle physics model of Dark Matter-CR interactions from experimental data on this scenario. 

\end{abstract}

\maketitle

{\bf \emph{Introduction-}}~~The existence of a non-luminous class of matter dubbed Dark Matter (DM) has been firmly established, albeit only on the basis of its gravitational impact on visible matter. Further, the particle nature of this DM is unknown. A non-gravitational detection of DM would provide an enormous first step in understanding its particle nature. 

One of the most promising experimental avenues is to search for the small energy depositions from DM elastically scattering in very sensitive detectors on Earth~\cite{Goodman:1984dc}. This ``direct detection'' of DM can proceed from scattering on either nuclei~\cite{Liu:2019kzq,Ren:2018gyx,Agnes:2018ves,Akerib:2017kat,Agnese:2017njq,Agnese:2017jvy,Angloher:2017sxg,Petricca:2017zdp,Aguilar-Arevalo:2016ndq,Amole:2017dex} or electrons~\cite{Agnese:2018col,Crisler:2018gci}. In either case however, the same interactions with ordinary matter allow high energy cosmic rays (CRs) to scatter on background DM. This has two important phenomenological impacts with observable consequences: (1) it can lead to additional CR energy losses~\cite{Cappiello:2018hsu}, and (2) can improve detection prospects for light DM by giving such particles much larger energies so that that are more easily detected by terrestrial nuclear~\cite{Bringmann:2018cvk,Cappiello:2019qsw} or electron scattering~\cite{Ema:2018bih,Cappiello:2019qsw}. Other recent methods for improving the bounds on light DM provide complementary probes using e.g. the Migdal effect~\cite{Ibe:2017yqa,Dolan:2017xbu,Akerib:2018hck,Armengaud:2019kfj,Bell:2019egg,Liu:2019kzq}, bremsstrahlung~\cite{Kouvaris:2016afs}, CMB distortions~\cite{Gluscevic:2017ywp,Boddy:2018kfv,Boddy:2018wzy}, and inelastic CR collision with the atmosphere producing energetic light dark matter via meson decays \cite{Alvey:2019zaa}. We note as well that relativistic scattering at DM direct detection experiments was previously considered in the context of models in which the products of DM annihilation scatter with nuclei~\cite{Cherry:2015oca}.

\begin{figure}[b!]
\includegraphics[angle=0,width=.4\textwidth]{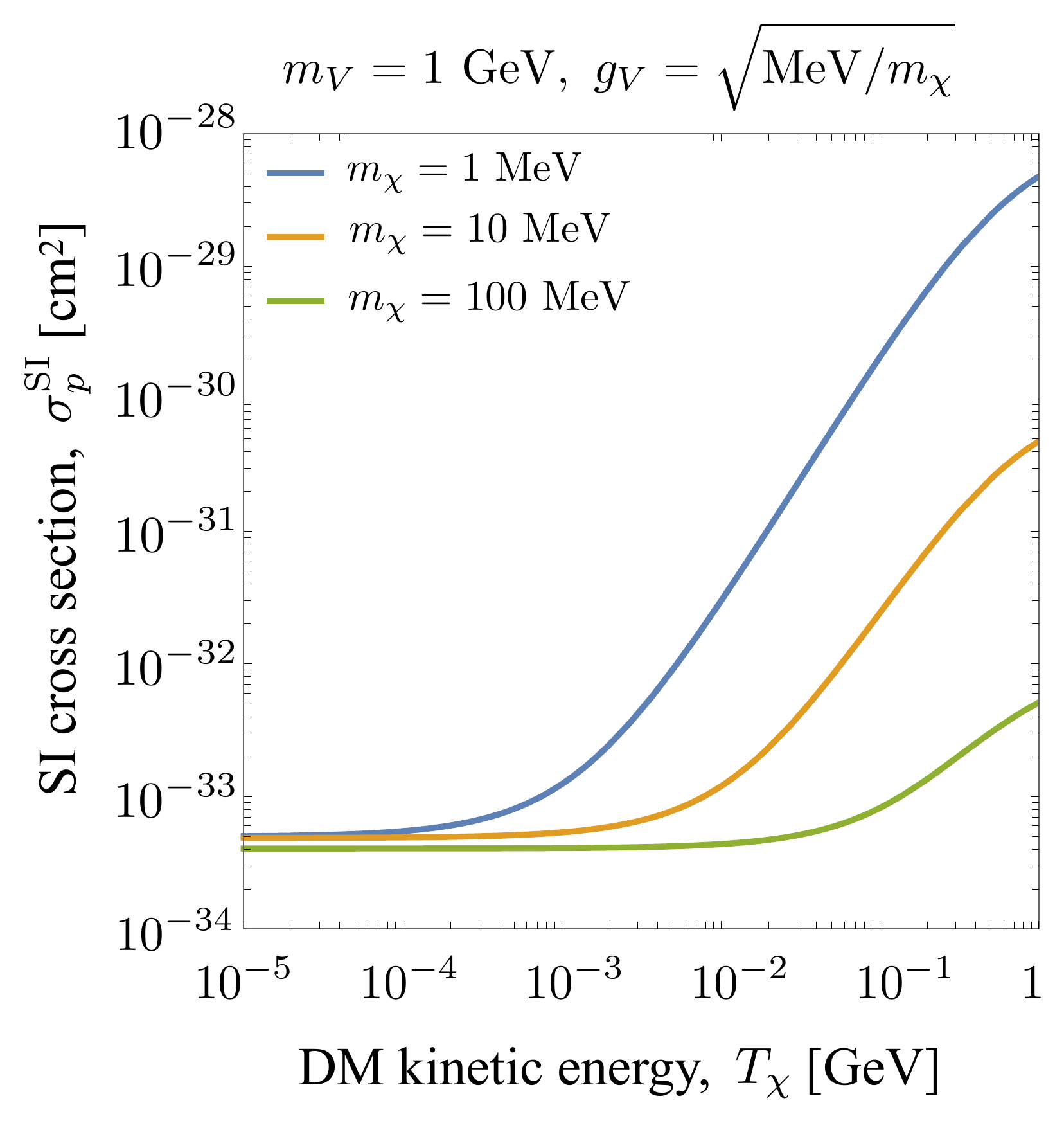}
\caption{Direct detection proton-DM scattering cross section (Eq.~\ref{eq:CRDMss}) as a function of incoming DM kinetic energy for a variety of DM masses. For illustration, the vector mediator mass is chosen to be $m_{V}$ = 1 GeV. The coupling choice $g = \sqrt{{\rm MeV}/m_{\chi}}$ ensures that the cross sections converge in the non-relativistic limit.}
\label{fig:cx}
\end{figure}


The existing literature on DM-CR interactions have assumed cross sections that are energy independent. This turns out to be a common occurrence for non-relativistic DM, but often not the case more generally. More typically, in elementary particle physics models in which the CR-DM interaction proceeds via the exchange of some new scalar or vector particle, the relativistic differential cross section depends non-trivially on the energy and mass of the participants. Moreover, DM-CR interactions probe large cross sections and can have large momentum transfers. If these large cross sections are generated by a mediating particle with mass $m_\phi^2 \lesssim q^2$, where $q$ is the exchanged momentum, one must retain the full propagator term in the cross section. Therefore to accurately formulate this problem, one must make use of a more complete model of the DM-CR interactions.

\begin{figure*}[t!]
\includegraphics[width=.46\textwidth]{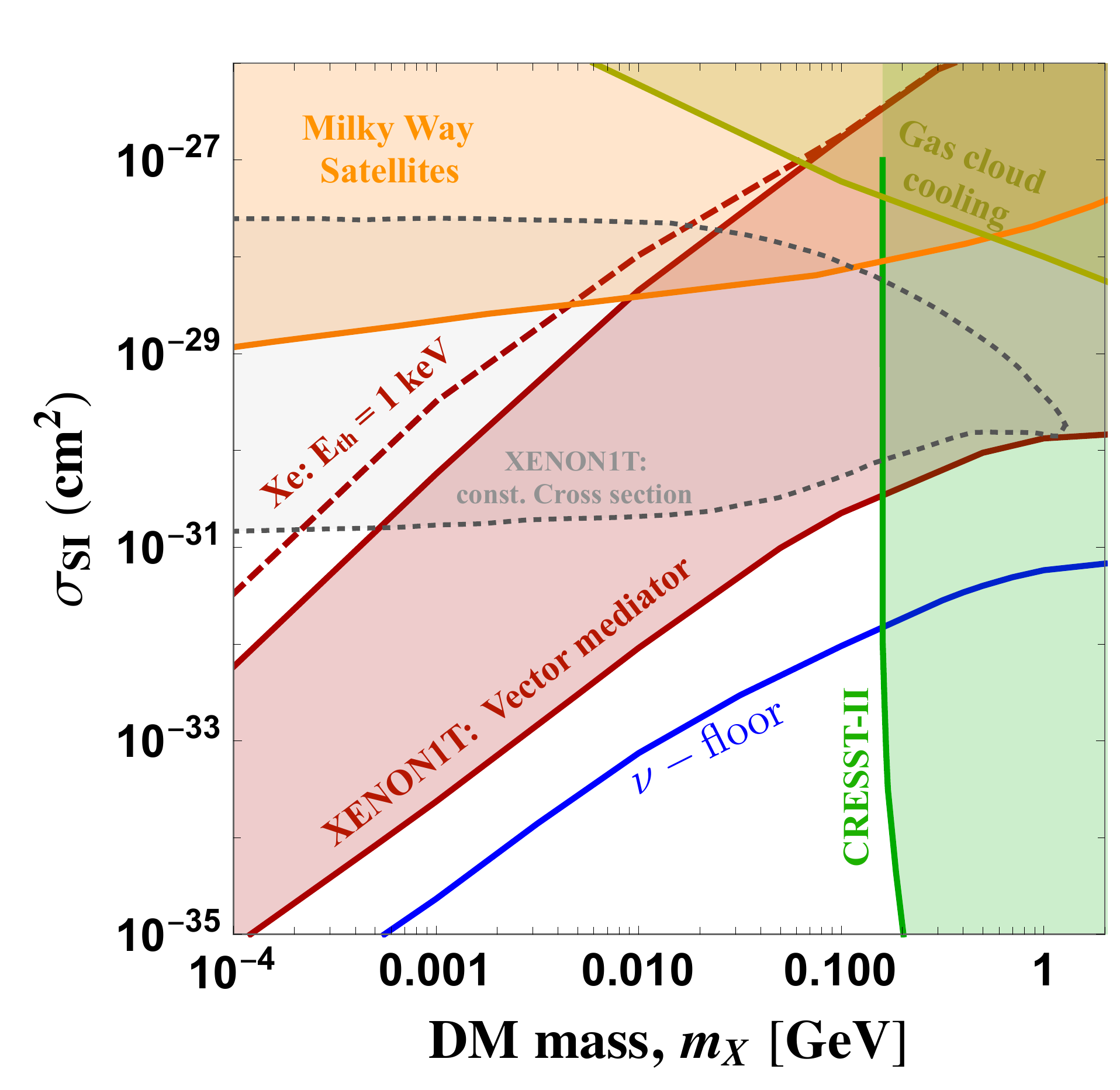}
\includegraphics[width=.46\textwidth]{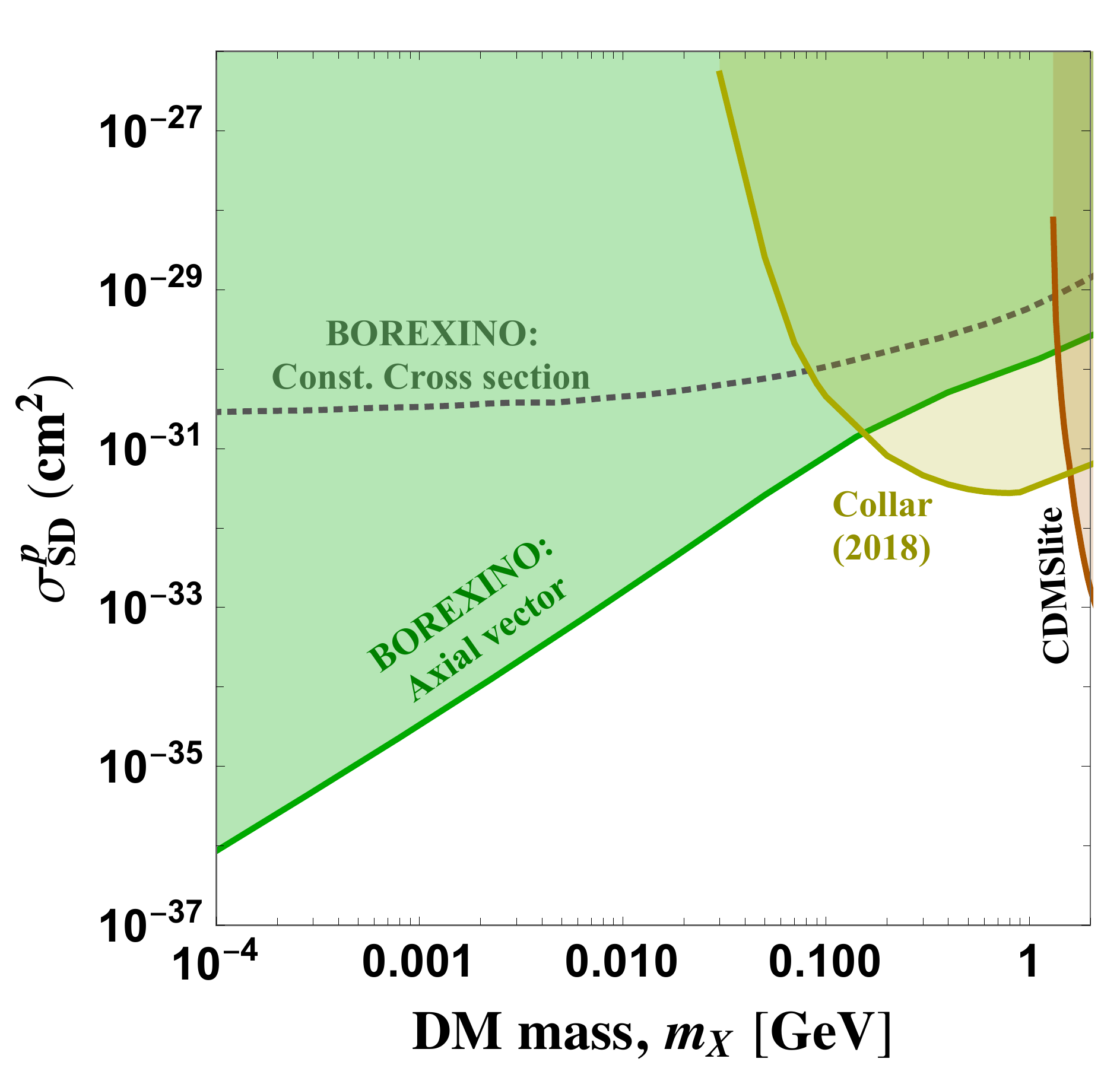}
\caption{Bounds on the spin-independent WIMP-proton cross section ({\it left panel}) and spin-dependepent ({\it right panel}) derived for the axial-vector mediator differential cross section. For vector interactions we examine XENON1T data (solid red line, $E_{\rm{th}} = 5~\rm{keV}$), and for axial vector interactions we use Borexino data. In both cases we have fixed the coupling, $g=1$. In the vector case, the upper dashed line ($E_{\rm{th}} = 1~\rm{keV}$) shows the projected sensitivity to dark matter attenuated by the Earth due to a lowered detector threshold. For comparison, bounds from \cite{Bringmann:2018cvk} which use an energy independent form of cross section are shown in dotted blue (left) and dotted green (right). We also display bounds on non-relativistic DM from the conventional Maxwell-Boltzmann distribution of local DM. The additional bounds we display include gas cloud cooling~\cite{Bhoonah:2018wmw}, Milky Way satellites~\cite{Nadler:2019zrb}, above ground CRESST-II~\cite{Angloher:2017sxg}, CDMSlite~\cite{Agnese:2017jvy}, and Collar~\cite{Collar:2018ydf}.} 
\label{fig:siBounds}
\end{figure*}

Here we formulate the DM-CR interactions within the context of a simplified model of DM-nucleon interactions(e.g.,~\cite{Pospelov:2007mp,Hooper:2008im,Cheung:2009qd,Essig:2010ye,Essig:2013lka,Dutta:2019fxn}). This allows us to remain agnostic of the detailed UV-physics, yet still capture the relativistic behavior of the scattering. This formalism is commonly applied in dark matter searches using collider~\cite{Buchmueller:2014yoa,Abdallah:2015ter}, indirect detection~\cite{Berlin:2014tja} and direct detection approaches~\cite{Dent:2015zpa}.  We will consider simplified models of a light vector, $V_\mu$,  and axial mediator, $A_\mu$, allowing us to cover the standard cases of spin-independent and spin-dependent scattering. 

The interaction Lagrangian takes the form, 
\bea
\mathscr{L}_{\rm int} &\supset& g_{\chi v} V_\mu \bar{\chi}\gamma^\mu \chi + g_{Nv} V_\mu \bar{N}\gamma^\mu N \nn
& & +g_{\chi a} A_\mu \bar{\chi}\gamma^\mu\gamma^5 \chi + g_{Na} A_\mu \bar{N}\gamma^\mu\gamma^5 N
\eea
where $\chi$ refers to the DM, and $N$ represents neutrons ($n$) and protons ($p$). We have also introduced the DM and nucleon couplings to the mediators $g_{\chi,i}$ and $g_{N,i}$, where $i=v,a$ denotes the vector and axial couplings, respectively. These interactions generate the differential cross sections 
\bea
\label{eq:CRDMss}
&&\left(\frac{d\sigma_{\chi N}}{dT_\chi}\right)_{\rm{vector},\rm{CR}} 
 =g_{\chi v}^{2}g_{N v}^{2}A^2F^2(q^2)
 \\\nonumber
 &\times&\frac{(m_\chi(m_N+T_i)^2-T_\chi((m_N+m_\chi)^2+2 m_\chi T_i)+m_\chi T^2_\chi)}{4\pi(2m_\chi T_\chi + m_v^2)^2(T_i^2 + 2m_T T_i)}
\eea
and
\bea
\label{eq:CRDMaa}
&&\left(\frac{d\sigma_{\chi N}}{dT_\chi}\right)_{\rm{axial},\rm{CR}}  \\ \nonumber
 &=&g_{\chi a}^{2}g_{Na}^{2}\frac{(4m_\chi m_N^2 +2T_\chi(m_\chi^2+m_N^2)+m_\chi T_\chi^2)}{8\pi(2m_\chi T_\chi + m_a^2)^2(T_i^2 + 2m_iT_i)},
\eea
where $A$ is the total nucleon number of the cosmic ray nucleus of mass $m_{\rm T}$, and $m_{\chi},m_N, m_i, m_{v}, m_a$ are the DM, nucleon, incident CR, vector and axial mediator masses, while $T_{\chi}$ is the outgoing DM kinetic energy. The dark matter is treated as being at rest in the galaxy (a good approximation for the cosmic ray energies under consideration). While the couplings may in principle be different, throughout this analysis we fix $g \equiv g_{\chi,i} = g_{N,i}$ for simplicity. For the form factor, $F^2(q^2)$, we have adopted the Helm model \cite{Helm:1956zz} for nuclei larger than hydrogen.

For the scattering from protons we follow \cite{Bringmann:2018cvk} and use the dipole form of \cite{Perdrisat:2006hj}. However, for heavier nuclei one needs to account for the momentum of the individual nucleons participating in the scattering process within the nucleus. For non-relativistic dark matter-nucleus scattering one can apply the standard treatment of \cite{Anand:2013yka}. In this work we adopt the recent analysis of neutrino-nucleus scattering analyzed in \cite{Bednyakov:2018mjd}, which closely resembles the physical situation of cosmic-ray scattered dark matter due to the relativistic nature of the incident neutrinos. This approach allows us to connect the nucleon cross-section in Eq.(\ref{eq:CRDMss}) to the full nuclear cross-section by relating the nucleon momenta to the full momentum transferred in the scattering process from energy-momentum conservation. This, for example, produces the propagator as a function of $m_T$ in the denominator of Eq.(\ref{eq:CRDMss}) rather than $m_N$ as the momentum is transferred to the full nucleus. 

In Fig.~\ref{fig:cx} we plot Eq.~\ref{eq:DD} for a variety of DM masses. The coupling choice $g = \sqrt{{\rm MeV}/m_{X}}$ ensures that the cross sections converge in the non-relativistic limit. As can be seen, there is a strong dependence on the DM mass, with light DM ($m_{\chi} < 1$ MeV) masses the cross section grows substantially with increasing CR energy, while for heavier DM ($m_{\chi} > 50$ MeV), the cross section drops precipitously. 

This simple result motivates a careful re-examination of the sensitivity of direct detection experiments to this relativistic flux of DM. We compute the bounds from direct detection focusing on the XENON1T experiment~\cite{Aprile:2018dbl}.




{\bf \emph{Calculational Framework- }}~~Here we reformulate the cosmic ray dark matter spectrum retaining energy dependence of the cross section, which is required for even the simplest vector-vector interaction (due to the relativistic nature of the scattering). The double-differential collision rate of CR particles $i$ with dark matter within an infinitesimal volume $dV$ is~\cite{Bringmann:2018cvk}
\bea
\frac{d^2\Gamma_{\rm{CR}_i\rightarrow\chi}}{dT_idT_\chi} = \frac{\rho_\chi}{m_\chi} \frac{d\sigma_{\chi i}}{dT_\chi} \frac{d\Phi^{\rm{LIS}}_i}{dT_i} dV
\eea
The scattered dark matter flux is then obtained by integrating this over the relevant volume and cosmic ray energies
\bea
\frac{d\Phi_\chi}{dT_\chi} &=& \int_V\!dV\int_{T_i^{\rm{min}}}\!\!\!\!dT_i \,\,\frac{d^2\Gamma_{\rm{CR}_i\rightarrow\chi}}{dT_idT_\chi}\\
&=& D_{\rm eff}\,\frac{\rho_\chi}{m_\chi}\, \sum_i \int_{T_i^{\rm{min}}}\!\!\!\!dT_i\,\frac{d\sigma_{\chi i}}{dT_\chi}\,\frac{d\Phi^{\rm{LIS}}_i}{dT_i} 
\eea
where $D_{{\rm eff}} = 1$ kpc is an effective diffusion zone parameter, $\rho_{\chi} = 0.3~{\rm GeV}{\rm cm}^{-3}$ is the local DM density, and $d\Phi^{\rm{LIS}}_i/dT_i$ is the local interstellar (LIS) flux of nuclear species $i$. 
Note that the parameter $D_{{\rm eff}}$ is essentially the distance over which the calculation accounts for CRs as the source of high-energy DM flux. While there is uncertainty in the precise value of this parameter, we follow Ref.~\cite{Bringmann:2018cvk} and conservatively set $D_{{\rm eff}} = 1$ kpc throughout.

Finally we compute the differential event rate (per unit detector mass) from the incoming relativistic DM flux in a direct detection experiment via
\be 
\frac{dR}{dE_{T}} =\frac{1}{m_{\rm T}} \int_{T_{\chi}^{{\rm min}}}^{\infty} dT_{\chi}~\frac{d\Phi_\chi}{dT_\chi}~ \frac{d \sigma_{\chi T}}{dE_{T}}
\label{eq:rate}
\ee
where $E_{T}$ is the target recoil energy. For vector and axial-vector exchange the total differential cross section $d\sigma_{\chi T}/dE_T$ is
obtained through the DM-$N$ cross-section 
\bea
&& \left(\frac{d\sigma_{\chi N}}{dE_{\rm{T}}}\right)_{{\rm vector},\rm{DD}} =g_{\chi v}^2g_{Nv}^2A^2m_{\rm T}[2(m_\chi+T_\chi)^2
\\\nonumber
&&-\frac{E_{\rm T}}{m_N^2}\left(m_{\rm T}m_{\chi}^2+m_N^2\left(m_{\rm T} + 2\left(m_\chi + T_\chi\right)\right)\right)+E_{\rm T}^2]
\\\nonumber
&&\times\left[\pi(2m_{\rm{T}}E_{\rm{T}} + m_v^2)^2(T_\chi^2 + 2m_\chi T_\chi)\right]^{-1}F^2(q^2)
\eea

\bea
&&\left(\frac{d\sigma}{dE_{\rm{T}}}\right)_{\rm{axial},\rm{DD}} = g_{\chi a}^2g_{Na}^2m_{\rm T}[2(m_\chi+T_\chi)^2
\\\nonumber
&&\times
-E_{\rm T}\left(m_{\rm T}\left(1+\left(\frac{m_\chi}{m_N}\right)^2\right)+2(m_\chi+T_\chi)-E_{\rm T}\right)]
\\\nonumber
&&\times[{\pi(2m_{\rm{T}}E_{\rm{T}} + m_a^2)^2(T_\chi^2 + 2m_\chi T_\chi)}]^{-1},
\label{eq:DD}
\eea
where $m_{\rm T}$ is the target mass of the detector nuclei.




{\bf \emph{Attenuation- }} Most direct detection experiments are deep underground in order to suppress large backgrounds from activity at the surface. For example, XENON1T is operated at the Laboratori Nazionali del Gran Sasso (LNGS) situated at a depth of $z\simeq 1.4$ km.  As a result of the Earth overburden, cross sections above some critical value will attenuate and decelerate the DM flux to such small energies that it is rendered undetectable (see e.g. Refs.~\cite{Collar:1993ss,Hasenbalg:1997hs,Zaharijas:2004jv,Mack:2007xj,Foot:2003iv,Kouvaris:2014lpa,Kavanagh:2016pyr,Emken:2017qmp,Emken:2018run,Bramante:2018qbc} for studies of the attenuation of non-relativistic DM). 

As in Ref.~\cite{Bringmann:2018cvk}, we evaluate the energy loss of dark matter of energy $T_\chi$ with respect to the distance $x$ traveled through the Earth. This depends on the energy $T_r$ lost by the dark matter in each collision through the relation
\bea
\label{eq:dTdx}
\frac{dT_\chi}{dx} = -\sum_T n_T \int \frac{d\sigma}{dT_r}T_rdT_r
\eea
where the sum is over the average nuclei densities $n_T$ of elements within the Earth. We use the values of $n_T$ provided by {\sf{DarkSUSY}} \cite{Bringmann:2018lay}.

While a detailed study of the effects of attenuation on CR-boosted DM is beyond the scope of this {\it letter}, we adopt the following prescription for determining the largest cross section underground experiments can probe. CRDM of a given mass will impinge upon the Earth with an initial energy, $T_{\chi,{\rm{in}}}$. As the dark matter travels through the Earth, it will lose energy as described in Eq.(\ref{eq:dTdx}) leaving it with a kinetic energy of $T_{\chi,z}$ once it reaches the depth, $z$, of XENON1T. For a given $T_{\chi,z}$, the maximal nuclear recoil within the detector is $T_i^{{\rm max}} = \frac{T_{\chi,z}^{2} + 2m_{\chi} T_{\chi,z}}{T_{\chi,z}+(m_{\chi}+m_{T_i})^{2}/2m_{T_i}}$. Thus, there will be a maximal cross-section above which CRDM striking the Earth with an energy $T_{\chi,\rm{in}}$, will be scattered to a kinetic energy below that which is necessary to generate a nuclear recoil above the detector threshold energy $E_{\rm{th}}$. As a simple prescription we determine the upper bounds of the cross section exclusion plot by finding the cross sections above which a $T_{\chi} =$ 1 GeV incoming state would attenuate below threshold. The present work only considers the effects from elastic scattering, and the upper bounds will change once inelastic effects are included. 

\begin{figure}[t!]
\includegraphics[width=.44\textwidth]{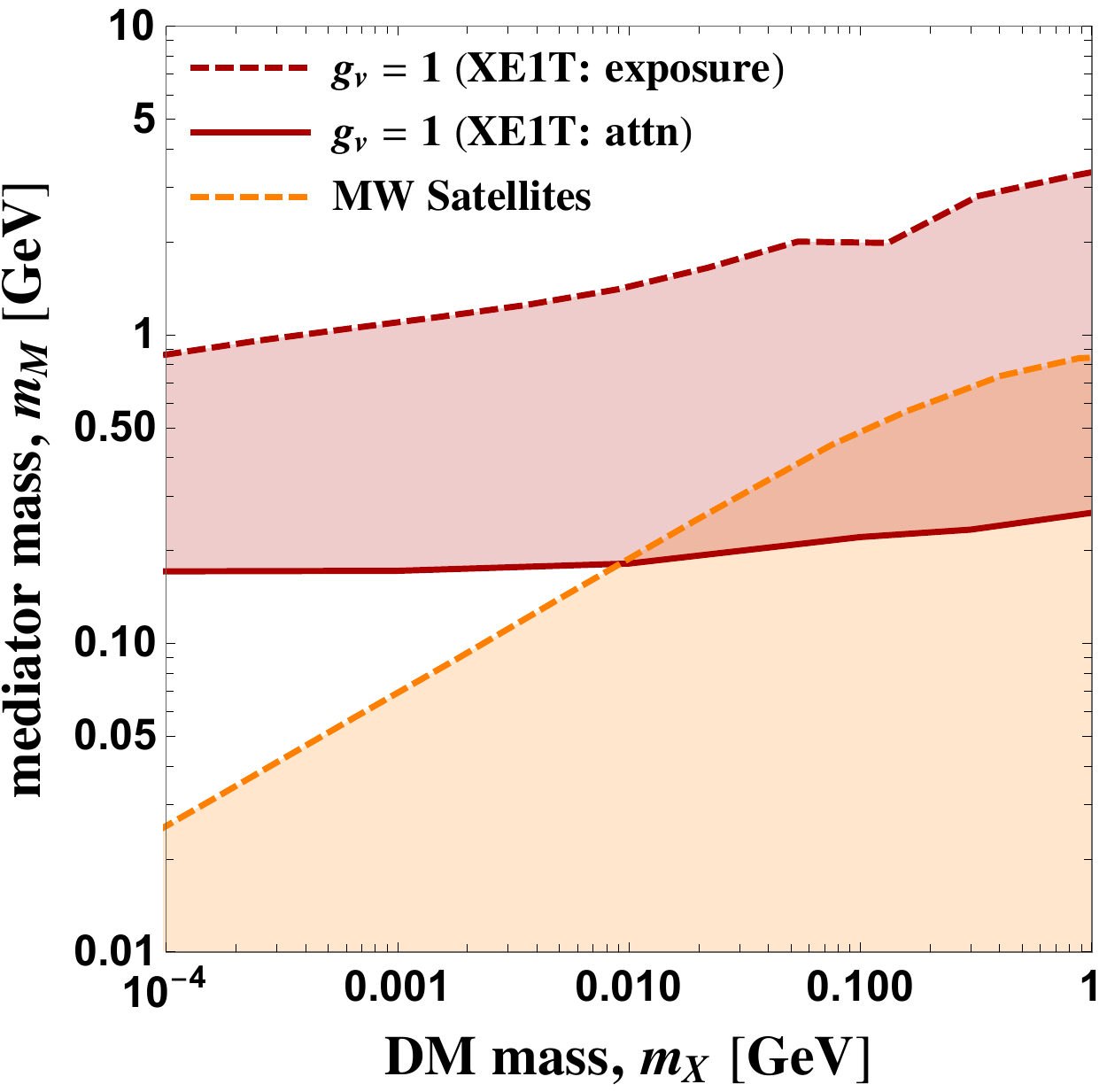}
\caption{Combination of bounds on the vector model with couplings set to unity. The red region shows the mediator and DM masses excluded by the search for  cosmic ray upscattered DM flux at XENON1T. The orange region is excluded by Milky Way satellties~\cite{Nadler:2019zrb}.}
\label{fig:med bounds}
\end{figure}


{\bf \emph{Results- }} In the two panels of Fig.~\ref{fig:siBounds} we present the results of our study by placing constraints on SI (using Xenon1T) and SD (using Borexino) cross-sections within our simplified model framework. To produce these bounds the coupling is fixed to $g=1$ and the mediator mass is varied as $\sigma=4g^4\mu^2/(\pi m_M^4)$. One can see that the exclusion region now strongly depends on the DM mass, with an exclusion region in the vector model case that is extended by orders of magnitude relative to an energy independent cross section as in Ref.~\cite{Bringmann:2018cvk}. {We also note that the upper bound on the cross-section displays a detector threshold dependence, emphasizing the enhanced reach of low threshold detection. 

{In Fig.~\ref{fig:siBounds} we do not display bounds from LHC constraints (also discussed in \cite{Cappiello:2018hsu,Cappiello:2019qsw}) where bounds from monojet plus missing energy searches \cite{Buchmueller:2014yoa} are no longer operable due to the large dark matter interactions within the detector, and instead give way to trackless hadronic energy deposition~\cite{Daci:2015hca}. LHC bounds may be relevant for some of the parameter space depending on the mediator mass explored in CRDM models (e.g., for $m_{M}\geq$ 1 GeV~\cite{Daci:2015hca}), but we leave this question for future work.} 

The neutrino background due to coherent elastic neutrino nucleus scattering (CE$\nu$NS) originating from solar and atmospheric neutrinos will provide an irreducible background to CRDM scattering in liquid xenon experiments. Since the nuclear recoil spectrum of CRDM differs from the CE$\nu$NS spectrum, there will be a true \textquoteleft neutrino floor'~\cite{Dent:2016iht}. To illustrate the sensitivity for which the neutrino background will start to become relevant, we calculate projected bounds for a XENON1t-like experiment with an exposure that expects to yield a single neutrino event (approximately 33 tonne-years). The curve denotes the approximate point at which the experiment's sensitivity to CRDM is altered from depending on the inverse square root of exposure, to the inverse fourth-root.  In the case of the spin-dependent scenario, Borexino is already sensitive to atmospheric neutrinos (see \cite{Atroshchenko_2016}) and thus can be considered to have already surpassed the floor.

It should be noted that in the two panels of Fig.~\ref{fig:siBounds}}, the constraints on the cross-section values are projections in the limit of no momentum transfer. This can obscure the bounds on the simplified model since there will be an interplay between the mediator mass and momentum transfer through the propagator. Therefore, it can be insightful to examine bounds in the, $m_M$ vs. $m_\chi$ plane, as shown in Fig.~\ref{fig:med bounds}. The coupling remains a free parameter and so we calculate the bounds for $g=1$ to show how the bounds depend on this choice. 
Larger couplings enable bounds on larger mediator masses. Conversely, lighter mediators enhance the rate and enable bounds with smaller couplings, but when $m_M^2<q^2$ the rate will not be enhanced further and thus $g\sim 0.1$ represents an approximate lower limit on the couplings accessible.  



{\bf \emph{Conclusions- }} In this brief {\it letter} we have considered the impact of realistic energy dependence on the flux and scattering rate of CR-boosted DM. Our work has a number of important consequences suggesting a number of avenues for future exploration. First, we have mostly focused on direct detection experiments, though as seen in the axial vector case, neutrino experiments can also probe relevant parameter space, especially given that many neutrino experiments are at shallower depths~\cite{Ema:2018bih,Bringmann:2018cvk}. These bounds also need to be revisited in light of the importance of energy dependent scattering.  A more complete treatment would include hadronic inelastic processes, which may be relevant for the initial upscattering, DM attenuation, and detector energy deposition. We note that inelastic processes in the dark sector have been shown to be relevant for increasing the DM upscattered flux in the initial CR-DM collision event~\cite{Alvey:2019zaa}

In addition, the energy dependence in simple, realistic models that we have explored here will also modify the constraints obtained in Ref.~\cite{Cappiello:2018hsu} from cosmic ray energy losses, as well as those obtained in \cite{Ema:2018bih} for a simplified model approach to DM-electron scattering when DM-electron couplings are introduced. For purposes of illustration we have focused on CR-DM interactions mediated by the exchange of a (axial-) vector mediator. This is of course only one possible class of models, and we plan to follow-up on this {\it letter} with an exploration of additional interactions mediating spin-independent and spin-dependent scattering including momentum dependent scattering. The CRDM bounds would be interesting for pseudo-scalar models where the usual direct detection bounds are very weak. We would also consider color mediator where the CR-DM bound would be much stronger due to the s-channel enhancement. Finally, due to the size of the energy transfer in the CRDM paradigm, it may be an intriguing space to explore inelastic dark matter models, as recently examined within boosted dark matter scenarios \cite{TuckerSmith:2001hy,Kim:2016zjx,Giudice:2017zke,Ha:2018obm}.~{Lastly we note that the CR-DM scattering may cease producing detectable signals for ultra-light DM masses.}



\vspace{1cm}

{\bf \emph{Acknowledgements-  }} We are grateful for helpful discussion with Chris Cappiello, Shunsaku Horiuchi, and Maxim Pospelov. BD acknowledge support from DOE Grant de-sc0010813. The work of I.M.S. is supported by the U.S. Department of Energy under the award number DE-SC0019163. JBD acknowledges support from the National Science Foundation under Grant No. NSF PHY-1820801. The research of JBD was also supported in part by NSF Grant No. PHY-1748958.

\bibliography{crdm}

\end{document}